\documentclass[10pt, a4paper, twocolumn, showpacs, titlepage, showkeys, aps]{revtex4-1}
\usepackage{times}
\usepackage{calligra}
\usepackage{hyperref}
\usepackage{amsmath}
\usepackage{amssymb}
\usepackage{graphicx}
\usepackage{subfigure}
\usepackage{booktabs}
\usepackage{rotating}
\usepackage{longtable}
\usepackage{bm}
\makeatletter

\newcommand{\Rmnum}[1]{\expandafter\@slowromancap\romannumeral #1@}
\makeatother

\begin{document}

\title{Tetrapartite entanglement measures of W-Class in noninertial frames}
\author{Ariadna J. Torres-Arenas$^1$}
\email{wia-k@hotmail.com (Ariadna J. Torres-Arenas).}
\author{Edgar O. L\'opez-Z\'u\~niga$^{1, 2}$}
\email{E-mail address: edgarosv.lz@gmail.com (Edgar O. L\'opez-Z\'uniga).}
\author{J. Antonio Salda\~na-Herrera$^{1, 2}$}
\email{E-mail address: jasaldanah@gmail.com (J. Antonio Salda\~na-Herrera). }
\author{Qian Dong$^{1}$}
\email{E-mail address: dldongqian@gmail.com (Q. Dong). }
\author{Guo-Hua Sun$^3$}
\email{sunghdb@yahoo.com (Guo-Hua Sun).}
\author{Shi-Hai Dong$^1$}
\email[Corresponding author:]{dongsh2@yahoo.com (S. H. Dong). }
\affiliation{$^1$ Laboratorio de Informaci\'{o}n Cu\'{a}ntica, CIDETEC, Instituto Polit\'{e}cnico Nacional, UPALM, CDMX 07700, Mexico}
\affiliation{$^2$ Facultad de Ciencias F\'{i}sico Matem\'{a}ticas, Universidad Aut\'{o}noma de Nuevo Le\'{o}n, San Nicol\'as de los Garza NL, 66450, Mexico}
\affiliation{$^3$ Catedr\'{a}tica CONACyT, Centro de Investigaci\'{o}n en Computaci\'{o}n, Instituto Polit\'{e}cnico Nacional, UPALM, CDMX 07738, Mexico}

\pacs{03. 67. -a, 03. 67. Mn, 03. 65. Ud, 04. 70. Dy}
\keywords{W-Class, Tetrapartite entanglement, Dirac field, noninertial frames, nonuniform acceleration}

\begin{abstract}
We present the entanglement measures of a tetrapartite W-Class entangled system in noninertial frame, where the transformation between Minkowski and Rindler coordinates is applied. Two cases are considered. First, when one qubit has uniform acceleration whilst the other three remain stationary. Second, when two qubits have nonuniform accelerations and the others stay inertial. The $1-1$ tangle, $1-3$ tangle and whole entanglement measurements ($\pi_4$ and $\Pi_4$), are studied and illustrated with graphics through their dependency on the acceleration parameter $r_d$ for the first case and $r_c$ and $r_d$ for the second case. It is found that the Negativities ($1-1$ tangle and $1-3$ tangle) and $\pi$-tangle decrease when the acceleration parameter $r_{d}$ or in the second case $r_c$ and $r_d$ increase, remaining a nonzero entanglement in the majority of the results. This means that the system will be always entangled except for special cases. It is shown that only the $1-1$ tangle for the first case, vanishes at infinite accelerations, but for the second case the $1-1$ tangle disappears completely when $r>0.472473$. It is found an analytical expression for von Neumann information entropy of the system and we notice that it increases with the acceleration parameter.
\end{abstract}
\maketitle

\section{Introduction}
Entanglement plays an important role in quantum information as a resource for quantum teleportation, communication and cryptography \cite{Bennet1, Bennet2, Bouwmeester}.
Since relativity allows us to have a fundamental understanding of theoretical model, it is relevant to study entanglement in a noninertial frame as a physical property of multipartite states \cite{Peres}.

Fuentes-Schuller and Mann first studied the entanglement of two qubits in noninertial frames and showed that the maximal bipartite entanglement decreases when one of two observers is accelerated. When the acceleration increases, the entanglement decreases until they reach infinite acceleration reducing the state to a separable one \cite{Fuentes}.
In the case of tripartite state, examined by Hwang {\it et al.}, it has been shown that just as the bipartite case, the entanglement degrades when one of the observers is accelerated, but the accelerated observer entanglement does not completely vanish, even when the observer is moving with an infinite acceleration. Also, an interesting result that they found is that entanglement is observer dependent \cite{Hwang}.

Li {\it et al.} analyzed the entanglement of a tetrapartite GHZ-state from one to four accelerated observers with a uniform acceleration $a$ using $\pi_{4}$- and $\Pi_{4}$-tangles which are the whole entanglement calculations \cite{Yazhou}. It should be addressed that other relevant contributions have been made in tripartite entangled systems \cite{yao14,khan14,khan142,bruschi2012,martin11}.

Rindler coordinates are used to describe the viewpoint of uniformly accelerated observers. There are necessary two different sets of coordinates to map field states in Minkowski space time to Rindler coordinates, which  define two disconnected regions in Rindler space-time \cite{Takagi, Martin}.

In this work we will investigate the tetrapartite entanglement of the Dirac Fields when one observer is accelerated and when two observers are accelerated for the W-Class of 4-particle case. Let Alice, Bob, Charlie and David share a four mode W-Class entangled state when they are initially not moving. First, we suppose that David moves with a uniform acceleration with respect to Alice, Bob and Charlie. Second, we suppose that Charlie and David are moving with nonuniform acceleration. We compute the $\pi_{4}$-tangle and the $\Pi_{4}$-tangle as function of David's acceleration and Charlie's and David's acceleration, respectively. We show that, just as the bipartite and tripartite entanglement cases, $\pi$-tangle decreases with the increasing acceleration, but unlike the bipartite entanglement case, when one observer reaches infinite acceleration, entanglement does not completely vanish but remaining a nonzero $\pi_{4}$-tangle and $\Pi_{4}$-tangle. This result implies the possibility of quantum information processing even when David reaches the Rindler horizon. Also, we are concerned with the study of von Neumann Entropy to quantify this tetrapartite entangled system.

The rest of this work is organized as follows. In section II we give a description of the system and the considerations are made and we discuss the tetrapartite entanglement of the W-Class when one observer is accelerated. In section III we discuss the tetrapartite entanglement of the W-Class when two observers are accelerated. We make calculations on Negativities and show  whole entanglement measurements. We will study von Neumann Entropy in section IV. Finally, in section V some discussions and concluding remarks are given.

\section{Tetrapartite entanglement when one observer is accelerated}

As shown by Verstraete {\it et al.}, there are nine different ways to entangle 4 qubits \cite{Verstraete}. The vast majority of papers in this subject focus on two main states: The Greenberger-Horne-Zeilinger (GHZ) state and the W-Class state. The last one often gets less attention than the first one since its calculations frequently get much more complex. There has already been a lot of treatment in the matter of the contributions for one, two, three and four acceleration qubits, for an initial GHZ-state. So our attention in this work will be for an initial W-Class entangled state.

The W-Class entangled state we will be considering is composed by fermions, which in this case are 4 qubits with the name of Alice, Bob, Charlie and David, each of them living in a different Hilbert Space. Now, we will be considering the case where three of them, namely, Alice, Bob and Charlie, are stationary while David moves with a uniform acceleration.

A generalization for $N$ qubits of the W-Class state has the form \cite{Dur}:
\begin{eqnarray}
\label{w-class}
\left|W\right\rangle_{N}=\frac{1}{\sqrt{N}}\left|N-1, 1\right\rangle,
\end{eqnarray}
where $\left|N-1, 1\right\rangle$ is the state involving $N-1$ zeros and a one. In this work, we take $N=4$ and use the subscripts $A, B, C$ and $D$ to represent the fermions Alice, Bob, Charlie and David, respectively. With this, the W-Class entangled state ($N=4$) can be written as follows:
\begin{widetext}
\begin{eqnarray}\label{w-class}
|W\rangle=\frac{1}{2}\left[|1_{\hat{A}}0_{\hat{B}}0_{\hat{C}}0_{\hat{D}}\right\rangle
+\left|0_{\hat{A}}1_{\hat{B}}0_{\hat{C}}0_{\hat{D}}\right\rangle
+\left|0_{\hat{A}}0_{\hat{B}}1_{\hat{C}}0_{\hat{D}}\right\rangle +\left|0_{\hat{A}}0_{\hat{B}}0_{\hat{C}}1_{\hat{D}}\right\rangle].
\end{eqnarray}
\end{widetext}

As discussed above, the most suitable way to describe an entangled state in a noninertial frame is using the Rindler coordinates. By that, we need to make the transformation between both coordinated systems. For the fermion field, we use the following transformation \cite{Wang}:	
\begin{equation}\label{0M}
\left|0_{w_{i}}\right\rangle_M=\cos r_{i}\left|0_{w_i}\right\rangle_{I}\left|0_{w_{i}}\right\rangle_{II}+\sin r_{i}\left|1_{w_i}\right\rangle_{I}\left|1_{w_{i}}\right\rangle_{II}
\end{equation}
and
\begin{eqnarray}\label{1M}
\left|1_{w_i}\right\rangle_M=\left|1_{w_i}\right\rangle_{I}\left|0_{w_i}\right\rangle_{II}
\end{eqnarray} with $\cos\, r_i = (e^{-2\pi\omega_{i}c/a_i}+1)^{-1/2}$, being $a_i$ the acceleration of the $i$-th accelerated observer and $w_i$ its respective frequency. One has $r_{i} \in [0, \frac{\pi}{4}]$ for $a_{i}\in[0, \infty)$. The subscripts $I$ and $II$ in the kets represent the Rindler modes for regions $I$ and $II$ in Rindler space-time diagram.

Applying Eqs. (\ref{0M}) and (\ref{1M}) to our $\left|W\right\rangle$ state we obtain the following state:

\begin{equation}
\begin{split}
\left|W\right\rangle=\frac{1}{2}\Big[\sin r_d \left|0_{\hat{A}}, 0_{\hat{B}}, 1_{\hat{C}}, 1_{\hat{{D_I}}}, 1_{\hat{{D_{II}}}}\right\rangle +\\
\sin r_d \left|0_{\hat{A}}, 1_{\hat{B}}, 0_{\hat{C}}, 1_{\hat{{D_I}}}, 1_{\hat{{D_{II}}}}\right\rangle +\\ \sin r_d \left|1_{\hat{A}}, 0_{\hat{B}}, 0_{\hat{C}}, 1_{\hat{{D_I}}}, 1_{\hat{{D_{II}}}}\right\rangle +\\
\cos r_d \left|0_{\hat{A}}, 0_{\hat{B}}, 1_{\hat{C}}, 0_{\hat{{D_I}}}, 0_{\hat{{D_{II}}}}\right\rangle +\\ \cos r_d \left|0_{\hat{A}}, 1_{\hat{B}}, 0_{\hat{C}}, 0_{\hat{{D_I}}}, 0_{\hat{{D_{II}}}}\right\rangle +\\
\cos r_d \left|1_{\hat{A}}, 0_{\hat{B}}, 0_{\hat{C}}, 0_{\hat{{D_I}}}, 0_{\hat{{D_{II}}}}\right\rangle +\\ \left|0_{\hat{A}}, 0_{\hat{B}}, 0_{\hat{C}}, 1_{\hat{{D_I}}}, 0_{\hat{{D_{II}}}}\right\rangle \Big].
\end{split}
\end{equation}

Usually, region $II$ is considered as the corresponding anti-particle region in the same frame \cite{Socolovsky, Mikio}. When David moves with a uniform acceleration in Rindler region $I$, he is causally disconnected from region
$II$. This means that he has no access to field modes in this region $II$. Thus, the observer has to trace over this inaccessible region so that the $II$ region will not be considered.

\subsection{Negativity}
The main equation defined for Negativity was made for bipartite state. This allows us to measure the entanglement of the system and defined as
\begin{eqnarray}\label{Neggeneral}
N_{\kappa \xi}=||\rho_{\kappa \xi}^{T_{\kappa}}||-1,
\end{eqnarray}where $\kappa$ and $\xi$ represent any qubit, $T_{\kappa}$ denotes the partial transpose of $\rho_{\kappa \xi}$ with respect to $\kappa$, and $||M||=Tr\sqrt{M^{\dagger}M}$ denotes the trace norm of a matrix $M$.

The equation (\ref{Neggeneral}) represents the general form to measure the Negativity. If we want to extend the formula for four qubits, we have \cite{Yazhou, Oliveira}:
\begin{eqnarray}
N_{\kappa (\xi \o \zeta)}=||\rho_{\kappa \xi \o \zeta}^{T_{\kappa}}||-1,
\end{eqnarray}
\begin{eqnarray}
N_{\kappa(\xi \o)}=||\rho_{\kappa\xi \o}^{T_{\kappa}}||-1,
\end{eqnarray}
\begin{eqnarray}
N_{\kappa \xi}=||\rho_{\kappa \xi}^{T_{\kappa}}||-1,
\end{eqnarray}which describe the entanglements $1-3$ tangle, $1-2$ tangle and $1-1$ tangle, respectively.

Alternatively, we know that
\begin{eqnarray}
||M||-1=2\sum_{i=1}^{N}|\lambda_{M}^{(-)}|^{i},
\end{eqnarray}
where $\lambda_{M}^{(-)}$ are the negative eigenvalues of the matrix $M$. In this work, for the $1-3$ tangle we have
 \begin{eqnarray}\label{eq11}
N_{\kappa (\xi \o \zeta)}=2\sum_{i=1}^{N}|\lambda_{\rho_{\kappa (\xi \o \zeta)}}^{(-)}|^{i}
\end{eqnarray}
and, for the $1-1$ tangle
 \begin{eqnarray}\label{eq12}
N_{\kappa \xi}=2\sum_{i=1}^{N}|\lambda_{\rho_{\kappa \xi}}^{(-)}|^{i}.
\end{eqnarray}

In order to begin Negativity computations, we trace over the unaccessible Rindler modes in Region $II$, and then we proceed to calculate the density matrix with the following forms,
\begin{equation}\label{}
\begin{array}{l}
\rho_{ABCD_{I}}=\frac{1}{4}\\
\left(\begin{array}{cccccccccccccccc}
 0 & 0 & 0 & 0 & 0 & 0 & 0 & 0 & 0 & 0 & 0 & 0 & 0 & 0 & 0 & 0 \\
 0 & 1 & \delta  & 0 & \delta  & 0 & 0 & 0 & \delta  & 0 & 0 & 0 & 0 & 0 & 0 & 0 \\
 0 & \delta  & \delta ^2 & 0 & \delta ^2 & 0 & 0 & 0 & \delta ^2 & 0 & 0 & 0 & 0 & 0 & 0 & 0 \\
 0 & 0 & 0 & \beta ^2 & 0 & \beta ^2 & 0 & 0 & 0 & \beta ^2 & 0 & 0 & 0 & 0 & 0 & 0 \\
 0 & \delta  & \delta ^2 & 0 & \delta ^2 & 0 & 0 & 0 & \delta ^2 & 0 & 0 & 0 & 0 & 0 & 0 & 0 \\
 0 & 0 & 0 & \beta ^2 & 0 & \beta ^2 & 0 & 0 & 0 & \beta ^2 & 0 & 0 & 0 & 0 & 0 & 0 \\
 0 & 0 & 0 & 0 & 0 & 0 & 0 & 0 & 0 & 0 & 0 & 0 & 0 & 0 & 0 & 0 \\
 0 & 0 & 0 & 0 & 0 & 0 & 0 & 0 & 0 & 0 & 0 & 0 & 0 & 0 & 0 & 0 \\
 0 & \delta  & \delta ^2 & 0 & \delta ^2 & 0 & 0 & 0 & \delta ^2 & 0 & 0 & 0 & 0 & 0 & 0 & 0 \\
 0 & 0 & 0 & \beta ^2 & 0 & \beta ^2 & 0 & 0 & 0 & \beta ^2 & 0 & 0 & 0 & 0 & 0 & 0 \\
 0 & 0 & 0 & 0 & 0 & 0 & 0 & 0 & 0 & 0 & 0 & 0 & 0 & 0 & 0 & 0 \\
 0 & 0 & 0 & 0 & 0 & 0 & 0 & 0 & 0 & 0 & 0 & 0 & 0 & 0 & 0 & 0 \\
 0 & 0 & 0 & 0 & 0 & 0 & 0 & 0 & 0 & 0 & 0 & 0 & 0 & 0 & 0 & 0 \\
 0 & 0 & 0 & 0 & 0 & 0 & 0 & 0 & 0 & 0 & 0 & 0 & 0 & 0 & 0 & 0 \\
 0 & 0 & 0 & 0 & 0 & 0 & 0 & 0 & 0 & 0 & 0 & 0 & 0 & 0 & 0 & 0 \\
 0 & 0 & 0 & 0 & 0 & 0 & 0 & 0 & 0 & 0 & 0 & 0 & 0 & 0 & 0 & 0 \\
\end{array}
\right)
\end{array}
\end{equation}

\begin{equation}\label{}
\begin{array}{l}
\rho_{ABCD_{I}}^{T_{A}}=\frac{1}{4}\\
\left(\begin{array}{cccccccccccccccc}
 0 & 0 & 0 & 0 & 0 & 0 & 0 & 0 & 0 & \delta  & \delta ^2 & 0 & \delta ^2 & 0 & 0 & 0 \\
 0 & 1 & \delta  & 0 & \delta  & 0 & 0 & 0 & 0 & 0 & 0 & \beta ^2 & 0 & \beta ^2 & 0 & 0 \\
 0 & \delta  & \delta ^2 & 0 & \delta ^2 & 0 & 0 & 0 & 0 & 0 & 0 & 0 & 0 & 0 & 0 & 0 \\
 0 & 0 & 0 & \beta ^2 & 0 & \beta ^2 & 0 & 0 & 0 & 0 & 0 & 0 & 0 & 0 & 0 & 0 \\
 0 & \delta  & \delta ^2 & 0 & \delta ^2 & 0 & 0 & 0 & 0 & 0 & 0 & 0 & 0 & 0 & 0 & 0 \\
 0 & 0 & 0 & \beta ^2 & 0 & \beta ^2 & 0 & 0 & 0 & 0 & 0 & 0 & 0 & 0 & 0 & 0 \\
 0 & 0 & 0 & 0 & 0 & 0 & 0 & 0 & 0 & 0 & 0 & 0 & 0 & 0 & 0 & 0 \\
 0 & 0 & 0 & 0 & 0 & 0 & 0 & 0 & 0 & 0 & 0 & 0 & 0 & 0 & 0 & 0 \\
 0 & 0 & 0 & 0 & 0 & 0 & 0 & 0 & \delta ^2 & 0 & 0 & 0 & 0 & 0 & 0 & 0 \\
 \delta  & 0 & 0 & 0 & 0 & 0 & 0 & 0 & 0 & \beta ^2 & 0 & 0 & 0 & 0 & 0 & 0 \\
 \delta ^2 & 0 & 0 & 0 & 0 & 0 & 0 & 0 & 0 & 0 & 0 & 0 & 0 & 0 & 0 & 0 \\
 0 & \beta ^2 & 0 & 0 & 0 & 0 & 0 & 0 & 0 & 0 & 0 & 0 & 0 & 0 & 0 & 0 \\
 \delta ^2 & 0 & 0 & 0 & 0 & 0 & 0 & 0 & 0 & 0 & 0 & 0 & 0 & 0 & 0 & 0 \\
 0 & \beta ^2 & 0 & 0 & 0 & 0 & 0 & 0 & 0 & 0 & 0 & 0 & 0 & 0 & 0 & 0 \\
 0 & 0 & 0 & 0 & 0 & 0 & 0 & 0 & 0 & 0 & 0 & 0 & 0 & 0 & 0 & 0 \\
 0 & 0 & 0 & 0 & 0 & 0 & 0 & 0 & 0 & 0 & 0 & 0 & 0 & 0 & 0 & 0 \\
\end{array}
\right)
\end{array}
\end{equation}

\begin{equation}\label{}
\begin{array}{l}
\rho_{ABCD_{I}}^{T_{B}}=\frac{1}{4}\\
\left(\begin{array}{cccccccccccccccc}
 0 & 0 & 0 & 0 & 0 & \delta  & \delta ^2 & 0 & 0 & 0 & 0 & 0 & \delta ^2 & 0 & 0 & 0 \\
 0 & 1 & \delta  & 0 & 0 & 0 & 0 & \beta ^2 & \delta  & 0 & 0 & 0 & 0 & \beta ^2 & 0 & 0 \\
 0 & \delta  & \delta ^2 & 0 & 0 & 0 & 0 & 0 & \delta ^2 & 0 & 0 & 0 & 0 & 0 & 0 & 0 \\
 0 & 0 & 0 & \beta ^2 & 0 & 0 & 0 & 0 & 0 & \beta ^2 & 0 & 0 & 0 & 0 & 0 & 0 \\
 0 & 0 & 0 & 0 & \delta ^2 & 0 & 0 & 0 & 0 & 0 & 0 & 0 & 0 & 0 & 0 & 0 \\
 \delta  & 0 & 0 & 0 & 0 & \beta ^2 & 0 & 0 & 0 & 0 & 0 & 0 & 0 & 0 & 0 & 0 \\
 \delta ^2 & 0 & 0 & 0 & 0 & 0 & 0 & 0 & 0 & 0 & 0 & 0 & 0 & 0 & 0 & 0 \\
 0 & \beta ^2 & 0 & 0 & 0 & 0 & 0 & 0 & 0 & 0 & 0 & 0 & 0 & 0 & 0 & 0 \\
 0 & \delta  & \delta ^2 & 0 & 0 & 0 & 0 & 0 & \delta ^2 & 0 & 0 & 0 & 0 & 0 & 0 & 0 \\
 0 & 0 & 0 & \beta ^2 & 0 & 0 & 0 & 0 & 0 & \beta ^2 & 0 & 0 & 0 & 0 & 0 & 0 \\
 0 & 0 & 0 & 0 & 0 & 0 & 0 & 0 & 0 & 0 & 0 & 0 & 0 & 0 & 0 & 0 \\
 0 & 0 & 0 & 0 & 0 & 0 & 0 & 0 & 0 & 0 & 0 & 0 & 0 & 0 & 0 & 0 \\
 \delta ^2 & 0 & 0 & 0 & 0 & 0 & 0 & 0 & 0 & 0 & 0 & 0 & 0 & 0 & 0 & 0 \\
 0 & \beta ^2 & 0 & 0 & 0 & 0 & 0 & 0 & 0 & 0 & 0 & 0 & 0 & 0 & 0 & 0 \\
 0 & 0 & 0 & 0 & 0 & 0 & 0 & 0 & 0 & 0 & 0 & 0 & 0 & 0 & 0 & 0 \\
 0 & 0 & 0 & 0 & 0 & 0 & 0 & 0 & 0 & 0 & 0 & 0 & 0 & 0 & 0 & 0 \\
\end{array}
\right)
\end{array}
\end{equation}

\begin{equation}\label{}
\begin{array}{l}
\rho_{ABCD_{I}}^{T_{C}}=\frac{1}{4}\\
\left(\begin{array}{cccccccccccccccc}
 0 & 0 & 0 & \delta  & 0 & 0 & \delta ^2 & 0 & 0 & 0 & \delta ^2 & 0 & 0 & 0 & 0 & 0 \\
 0 & 1 & 0 & 0 & \delta  & 0 & 0 & \beta ^2 & \delta  & 0 & 0 & \beta ^2 & 0 & 0 & 0 & 0 \\
 0 & 0 & \delta ^2 & 0 & 0 & 0 & 0 & 0 & 0 & 0 & 0 & 0 & 0 & 0 & 0 & 0 \\
 \delta  & 0 & 0 & \beta ^2 & 0 & 0 & 0 & 0 & 0 & 0 & 0 & 0 & 0 & 0 & 0 & 0 \\
 0 & \delta  & 0 & 0 & \delta ^2 & 0 & 0 & 0 & \delta ^2 & 0 & 0 & 0 & 0 & 0 & 0 & 0 \\
 0 & 0 & 0 & 0 & 0 & \beta ^2 & 0 & 0 & 0 & \beta ^2 & 0 & 0 & 0 & 0 & 0 & 0 \\
 \delta ^2 & 0 & 0 & 0 & 0 & 0 & 0 & 0 & 0 & 0 & 0 & 0 & 0 & 0 & 0 & 0 \\
 0 & \beta ^2 & 0 & 0 & 0 & 0 & 0 & 0 & 0 & 0 & 0 & 0 & 0 & 0 & 0 & 0 \\
 0 & \delta  & 0 & 0 & \delta ^2 & 0 & 0 & 0 & \delta ^2 & 0 & 0 & 0 & 0 & 0 & 0 & 0 \\
 0 & 0 & 0 & 0 & 0 & \beta ^2 & 0 & 0 & 0 & \beta ^2 & 0 & 0 & 0 & 0 & 0 & 0 \\
 \delta ^2 & 0 & 0 & 0 & 0 & 0 & 0 & 0 & 0 & 0 & 0 & 0 & 0 & 0 & 0 & 0 \\
 0 & \beta ^2 & 0 & 0 & 0 & 0 & 0 & 0 & 0 & 0 & 0 & 0 & 0 & 0 & 0 & 0 \\
 0 & 0 & 0 & 0 & 0 & 0 & 0 & 0 & 0 & 0 & 0 & 0 & 0 & 0 & 0 & 0 \\
 0 & 0 & 0 & 0 & 0 & 0 & 0 & 0 & 0 & 0 & 0 & 0 & 0 & 0 & 0 & 0 \\
 0 & 0 & 0 & 0 & 0 & 0 & 0 & 0 & 0 & 0 & 0 & 0 & 0 & 0 & 0 & 0 \\
 0 & 0 & 0 & 0 & 0 & 0 & 0 & 0 & 0 & 0 & 0 & 0 & 0 & 0 & 0 & 0 \\
\end{array}
\right)
\end{array}
\end{equation}	

\begin{equation}\label{}
\begin{array}{l}
\rho_{ABCD_{I}}^{T_{D_{I}}}=\frac{1}{4}\\
\left(\begin{array}{cccccccccccccccc}
 0 & 0 & 0 & \delta  & 0 & \delta  & 0 & 0 & 0 & \delta  & 0 & 0 & 0 & 0 & 0 & 0 \\
 0 & 1 & 0 & 0 & 0 & 0 & 0 & 0 & 0 & 0 & 0 & 0 & 0 & 0 & 0 & 0 \\
 0 & 0 & \delta ^2 & 0 & \delta ^2 & 0 & 0 & 0 & \delta ^2 & 0 & 0 & 0 & 0 & 0 & 0 & 0 \\
 \delta  & 0 & 0 & \beta ^2 & 0 & \beta ^2 & 0 & 0 & 0 & \beta ^2 & 0 & 0 & 0 & 0 & 0 & 0 \\
 0 & 0 & \delta ^2 & 0 & \delta ^2 & 0 & 0 & 0 & \delta ^2 & 0 & 0 & 0 & 0 & 0 & 0 & 0 \\
 \delta  & 0 & 0 & \beta ^2 & 0 & \beta ^2 & 0 & 0 & 0 & \beta ^2 & 0 & 0 & 0 & 0 & 0 & 0 \\
 0 & 0 & 0 & 0 & 0 & 0 & 0 & 0 & 0 & 0 & 0 & 0 & 0 & 0 & 0 & 0 \\
 0 & 0 & 0 & 0 & 0 & 0 & 0 & 0 & 0 & 0 & 0 & 0 & 0 & 0 & 0 & 0 \\
 0 & 0 & \delta ^2 & 0 & \delta ^2 & 0 & 0 & 0 & \delta ^2 & 0 & 0 & 0 & 0 & 0 & 0 & 0 \\
 \delta  & 0 & 0 & \beta ^2 & 0 & \beta ^2 & 0 & 0 & 0 & \beta ^2 & 0 & 0 & 0 & 0 & 0 & 0 \\
 0 & 0 & 0 & 0 & 0 & 0 & 0 & 0 & 0 & 0 & 0 & 0 & 0 & 0 & 0 & 0 \\
 0 & 0 & 0 & 0 & 0 & 0 & 0 & 0 & 0 & 0 & 0 & 0 & 0 & 0 & 0 & 0 \\
 0 & 0 & 0 & 0 & 0 & 0 & 0 & 0 & 0 & 0 & 0 & 0 & 0 & 0 & 0 & 0 \\
 0 & 0 & 0 & 0 & 0 & 0 & 0 & 0 & 0 & 0 & 0 & 0 & 0 & 0 & 0 & 0 \\
 0 & 0 & 0 & 0 & 0 & 0 & 0 & 0 & 0 & 0 & 0 & 0 & 0 & 0 & 0 & 0 \\
 0 & 0 & 0 & 0 & 0 & 0 & 0 & 0 & 0 & 0 & 0 & 0 & 0 & 0 & 0 & 0 \\
\end{array}
\right)
\end{array}
\end{equation}
where and hereafter we will use the substitutions $\sin r_c \to \alpha , \sin r_d \to \beta , \cos r_c \to \gamma , \cos r_d \to \delta$.

We now proceed to calculate the Negativities, which will help us to find out the whole entanglement measurements. We are going to make use of equation (\ref{eq11}) to calculate the $1-3$ entanglement with the following form
\begin{equation}\label{1-3ABCone}
N_{A(BCD_{I})} = N_{B(ACD_{I})}=N_{C(ABD_{I})},
\end{equation}
\begin{equation}\label{1-3Done}
\begin{array}{l}
N_{D_{I}(ABC)}=\\
\frac{1}{8} \left[3 \cos{2r_d}+\sqrt{\frac{3}{2}} \sqrt{4 \cos{2r_d}+3 \cos {4r_d}+25}-3\right].
\end{array}
\end{equation}

As we can see in (\ref{1-3ABCone}) the measurements when the observer is stationary no matter it is Alice, Bob or Charlie are equal to each other (FIG.1(a)). Due to their complicated expressions we do not attempt to write them out explicitly. Fortunately, the expression for $N_{D_{I}(ABC)}$ is simpler and can be obtained analytically.

Now, we proceed to calculate the $1-1$ tangle using equation (\ref{eq12}) and obtain the corresponding density matrices by tracing over two of four qubits in the system respectively for each case.
Because of the symmetry among Alice, Bob and Charlie, we have the $1-1$ tangle expressed as a combination of pairs of the elements $\{$A, B, C$\}$
\begin{eqnarray}
\rho_{AB}^{T_{A}}=\rho_{CA}^{T_{C}}=\rho_{BC}^{T_{B}}=\frac{1}{4}\left(
\begin{array}{cccc}
2 & 0 & 0 & 1 \\
0 & 1 & 0 & 0 \\
0 & 0 & 1 & 0 \\
1 & 0 & 0 & 0 \\
\end{array}
\right)
\end{eqnarray}

On the other hand, due to symmetry the $1-1$ tangle among two sets  $i$=$\{$A, B, C$\}$ and $j$=$\{$ D$_I$$\}$ is expressed as
\begin{equation}
\rho_{ij}^{T_{i}}=\frac{1}{4}\left(
\begin{array}{cccc}
2\cos ^2{r_d}& 0 & 0 & \cos{r_d}\\
0 & 2 \sin ^2{r_d}+1& 0 & 0 \\
0 & 0 & \cos ^2{r_d}& 0 \\
\cos{r_d}& 0 & 0 & \sin ^2{r_d}\\
\end{array}
\right),
\end{equation}

\begin{eqnarray}
\label{1-1onecte}
N_{AB}=N_{BC}=N_{AC}=\frac{1}{2} \left(\sqrt{2}-1\right),
\end{eqnarray}

\begin{equation}
\label{1-1onevariant}
\begin{split}
N_{ij}=\frac{1}{16}
\left(-6+\sqrt{2} \sqrt{28 \cos {2r_d}+9 \cos {4r_d}+27}-2 \cos {2r_d}\right).
\end{split}
\end{equation}

We can see that $N_{AB}$ in (\ref{1-1onecte}) is a constant even in the infinite limit. On the other hand, the observed decrement $N_{ij}$ in (\ref{1-1onevariant}) is due to the $r_d$ parameter (FIG.1(b)). It is worth noting that while the 1-1 tangle vanishes at infinite acceleration for the accelerated observer it is not the case for the 1-3 tangle.
\begin{figure}[htbp]
\subfigure[]{\includegraphics[width=4cm]{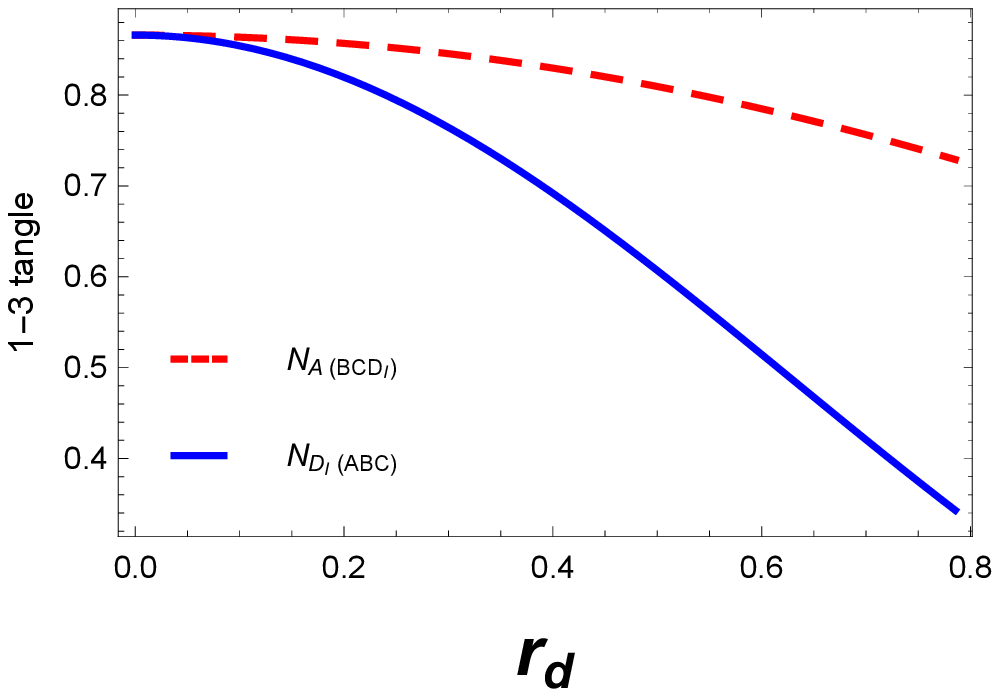}}\label{1-3 tangle}\hfil%
\subfigure[]{\includegraphics[width=4cm]{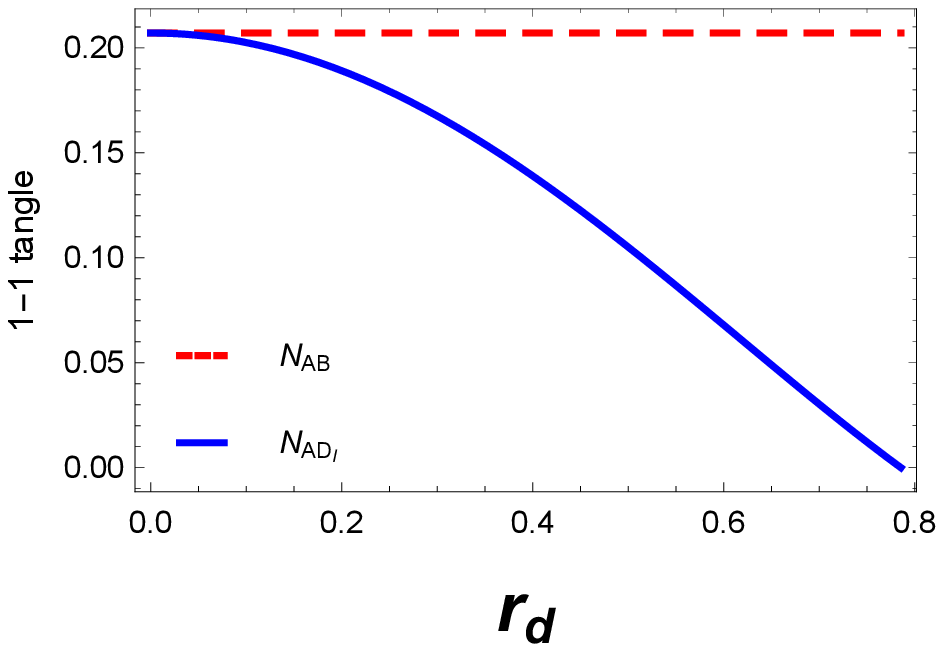}}\label{1-1 tangle}
\caption{\label{1-1 tangle}(a)  \text{$1-3$ tangle} of Alice $N_{A(BCD_I)}$  and David $N_{D_{I}(ABC)}$, respectively as a function of the acceleration parameter $r_d$, (b) $1-1$ tangle $N_{AB}$ and $N_{{ij}}$ as a function of the acceleration parameter $r_d$. }
\end{figure}

\subsection{The $\pi$-tangle entanglement measures }
Another important entanglement measurements are defined for tripartite states, which are the whole entanglement measurements ($\pi_3$ and $\Pi_3$). Both of them depend on the Negativities and we can make a tetrapartite extension of the latter equations yielding \cite{Oliveira}
\begin{eqnarray}\label{eq24-27}
\pi_{\kappa}=N_{\kappa(\xi \o \zeta)}^{2}-N_{\kappa \xi}^{2}-N_{\kappa \o}^{2}-N_{\kappa \zeta}^2,\\
\pi_{\xi}=N_{\xi(\kappa \o \zeta)}^{2}-N_{\xi\kappa}^{2}-N_{\xi \o}^{2}-N_{\xi \zeta}^2,\\
\pi_{\o}=N_{\o(\kappa \xi \zeta)}^{2}-N_{\o\kappa}^{2}-N_{\o\xi}^{2}-N_{\o \zeta}^2,\\
\pi_{\zeta}=N_{\zeta(\kappa \xi \o)}^{2}-N_{\zeta\kappa}^{2}-N_{\zeta\xi}^{2}-N_{\zeta \o}^2.
\end{eqnarray}

Now, we are able to obtain the $\pi_4$-tangle, which is given by taking the average of $\pi_{\kappa}$, $\pi_{\xi}$, $\pi_{\xi}$ and $\pi_{\zeta}$ \cite{Yazhou}
\begin{eqnarray}\label{eq28}
\pi_{4}=\frac{1}{4}\left(\pi_{\kappa}+\pi_{\xi}+\pi_{\o}+\pi_{\zeta}\right).
\end{eqnarray}

Also, we use another whole entanglement measure defined as \cite{Sabin}
\begin{eqnarray}\label{eq29}
\Pi_4=\left(\pi_{\kappa}+\pi_{\xi}+\pi_{\o}+\pi_{\zeta}\right)^{\frac{1}{4}}.
\end{eqnarray}

For the present case, one has
\begin{eqnarray}
\pi_{A}=N_{A(BCD_I)}^{2}-N_{AB}^{2}-N_{AC}^{2}-N_{AD_I}^2,\\
\pi_{B}=N_{B(ACD_I)}^{2}-N_{BA}^{2}-N_{BC}^{2}-N_{BD_I}^2,\\
\pi_{C}=N_{C(ABD_I)}^{2}-N_{CA}^{2}-N_{CB}^{2}-N_{CD_I}^2,\\
\pi_{D_I}=N_{D_I(ABC)}^{2}-N_{D_IA}^{2}-N_{D_IB}^{2}-N_{D_IC}^2,
\end{eqnarray}from which we are able to calculate
\begin{eqnarray}
\pi_{4}=\frac{1}{4}\left(\pi_A+\pi_B+\pi_C+\pi_{D_I}\right)
\end{eqnarray}
and
\begin{eqnarray}
\Pi_4=\left(\pi_A\pi_B\pi_C\pi_{D_I}\right)^{\frac{1}{4}}.
\end{eqnarray}

\begin{figure}[htbp]{\includegraphics[width=6cm]{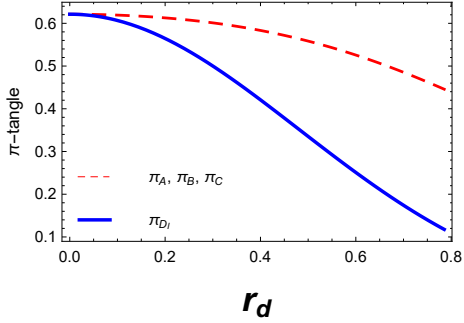}}\label{Residual entanglement}
\caption{\label{Residual entanglement} \text{Residual entanglement} of Alice $\pi_A$ (Bob and Charlie)  and David $\pi_DI$, respectively as a function of the acceleration parameter $r_d$. }
\end{figure}
\begin{figure}[htbp]
\includegraphics[width=6cm]{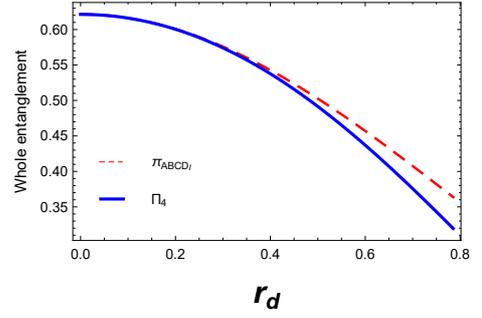}\label{Pi-tangle}
\caption{\label{Pi-tangle} \text{$\pi_{4}$-tangle} as a function of the acceleration parameter $r_d$, $\Pi_4$ whole entanglement as a function of the acceleration parameter $r_d$. }
\end{figure}

For this case we are studying, we have $\pi_A=\pi_B=\pi_C>\pi_{D_I}$. We plot $\pi_A$ ($\pi_B$, $\pi_C$) and $\pi_{D_I}$ residual entanglements in FIG. 2. and $\pi_{ABCD}$-tangle and $\Pi_4$ whole entanglements in FIG.3.

\section{Tetrapartite entanglement when two observers are accelerated}

In this case, similar to above approach we are about to study the tetrapartite entanglement measurements of the system when two observers are accelerated. Computations become more complex due to the multi-qubit state. We are working with the double Rindler transformations in the state, leading to a bigger dependence of trigonometric functions. Now, we suppose that Charlie and David are the accelerated observers. Applying equations (\ref{0M}) and (\ref{1M}) to our initial $|W\rangle$ state, we obtain:
\begin{widetext}
\begin{equation}
\begin{split}
\left|W\right\rangle=\frac{1}{2}\Big( \sin r_c\sin r_d \left|0_{\hat{A}}, 1_{\hat{B}}, 1_{\hat{C_I}}, 1_{\hat{C_{II}}}, 1_{\hat{{D_I}}}, 1_{\hat{{D_{II}}}}\right\rangle +
 \sin r_c\sin r_d \left|1_{\hat{A}}, 0_{\hat{B}}, 1_{\hat{C_I}}, 1_{\hat{C_{II}}}, 1_{\hat{{D_I}}}, 1_{\hat{{D_{II}}}}\right\rangle +\\ \cos r_c\cos r_d \left|0_{\hat{A}}, 1_{\hat{B}}, 0_{\hat{C_I}}, 0_{\hat{C_{II}}}, 0_{\hat{{D_I}}}, 0_{\hat{{D_{II}}}}\right\rangle +
  \cos r_c\cos r_d \left|1_{\hat{A}}, 0_{\hat{B}}, 0_{\hat{C_I}}, 0_{\hat{C_{II}}}, 0_{\hat{{D_I}}}, 0_{\hat{{D_{II}}}}\right\rangle+\\ \cos r_c\sin r_d \left|0_{\hat{A}}, 1_{\hat{B}}, 0_{\hat{C_I}}, 0_{\hat{C_{II}}}, 1_{\hat{{D_I}}}, 1_{\hat{{D_{II}}}}\right\rangle+
   \sin r_c \cos r_d\left|0_{\hat{A}}, 1_{\hat{B}}, 1_{\hat{C_I}}, 1_{\hat{C_{II}}}, 0_{\hat{{D_I}}}, 0_{\hat{{D_{II}}}}\right\rangle+\\ \cos r_c\sin r_d \left|1_{\hat{A}}, 0_{\hat{B}}, 0_{\hat{C_I}}, 0_{\hat{C_{II}}}, 1_{\hat{{D_I}}}, 1_{\hat{{D_{II}}}}\right\rangle+ \sin r_c\cos r_d \left|1_{\hat{A}}, 0_{\hat{B}}, 1_{\hat{C_I}}, 1_{\hat{C_{II}}}, 0_{\hat{{D_I}}}, 0_{\hat{{D_{II}}}}\right\rangle+\\ \sin r_c\left|0_{\hat{A}}, 0_{\hat{B}}, 1_{\hat{C_I}}, 1_{\hat{C_{II}}}, 1_{\hat{{D_I}}}, 0_{\hat{{D_{II}}}}\right\rangle+
  \cos r_c \left|0_{\hat{A}}, 0_{\hat{B}}, 0_{\hat{C_I}}, 0_{\hat{C_{II}}}, 1_{\hat{{D_I}}}, 0_{\hat{{D_{II}}}}\right\rangle+\\ \sin r_d \left|0_{\hat{A}}, 0_{\hat{B}}, 1_{\hat{C_I}}, 0_{\hat{C_{II}}}, 1_{\hat{{D_I}}}, 1_{\hat{{D_{II}}}}\right\rangle+ \cos r_d \left|0_{\hat{A}}, 0_{\hat{B}}, 1_{\hat{C_I}}, 0_{\hat{C_{II}}}, 0_{\hat{{D_I}}}, 0_{\hat{{D_{II}}}}\right\rangle\Big),
\end{split}
\end{equation}
\end{widetext}
where $r_c$ and $r_d$ are the acceleration parameters for Charlie and David, respectively. Notations $C_I$, $C_{II}$, $D_I$ and $D_{II}$ represent that Charlie and David move with an acceleration in Rindler Regions $I$ and $II$. Just as the previous calculations, we are not going to consider region $II$.

\subsection{Negativity}

We will use one more time equations (\ref{eq11}) and (\ref{eq12}) to determine the degree of entanglement for the 1-3 and 1-1 tangle. After tracing over the unaccessible Rindler modes $C_{II}$ and $D_{II}$ we obtain the density matrix shown as
\begin{widetext}
\begin{equation}\label{eq37}
\rho_{ABC_{I}D_{I}}=\frac{1}{4}\left(
\begin{array}{cccccccccccccccc}
 0 & 0 & 0 & 0 & 0 & 0 & 0 & 0 & 0 & 0 & 0 & 0 & 0 & 0 & 0 & 0 \\
 0 & \gamma ^2 & \gamma  \delta  & 0 & \gamma ^2 \delta  & 0 & 0 & 0 & \gamma ^2 \delta  & 0 & 0 & 0 & 0 & 0 & 0 & 0 \\
 0 & \gamma  \delta  & \delta ^2 & 0 & \gamma  \delta ^2 & 0 & 0 & 0 & \gamma  \delta ^2 & 0 & 0 & 0 & 0 & 0 & 0 & 0 \\
 0 & 0 & 0 & \alpha ^2+\beta ^2 & 0 & \beta ^2 \gamma  & \alpha ^2 \delta  & 0 & 0 & \beta ^2 \gamma  & \alpha ^2 \delta  & 0 & 0 & 0 & 0 & 0 \\
 0 & \gamma ^2 \delta  & \gamma  \delta ^2 & 0 & \gamma ^2 \delta ^2 & 0 & 0 & 0 & \gamma ^2 \delta ^2 & 0 & 0 & 0 & 0 & 0 & 0 & 0 \\
 0 & 0 & 0 & \beta ^2 \gamma  & 0 & \beta ^2 \gamma ^2 & 0 & 0 & 0 & \beta ^2 \gamma ^2 & 0 & 0 & 0 & 0 & 0 & 0 \\
 0 & 0 & 0 & \alpha ^2 \delta  & 0 & 0 & \alpha ^2 \delta ^2 & 0 & 0 & 0 & \alpha ^2 \delta ^2 & 0 & 0 & 0 & 0 & 0 \\
 0 & 0 & 0 & 0 & 0 & 0 & 0 & \alpha ^2 \beta ^2 & 0 & 0 & 0 & \alpha ^2 \beta ^2 & 0 & 0 & 0 & 0 \\
 0 & \gamma ^2 \delta  & \gamma  \delta ^2 & 0 & \gamma ^2 \delta ^2 & 0 & 0 & 0 & \gamma ^2 \delta ^2 & 0 & 0 & 0 & 0 & 0 & 0 & 0 \\
 0 & 0 & 0 & \beta ^2 \gamma  & 0 & \beta ^2 \gamma ^2 & 0 & 0 & 0 & \beta ^2 \gamma ^2 & 0 & 0 & 0 & 0 & 0 & 0 \\
 0 & 0 & 0 & \alpha ^2 \delta  & 0 & 0 & \alpha ^2 \delta ^2 & 0 & 0 & 0 & \alpha ^2 \delta ^2 & 0 & 0 & 0 & 0 & 0 \\
 0 & 0 & 0 & 0 & 0 & 0 & 0 & \alpha ^2 \beta ^2 & 0 & 0 & 0 & \alpha ^2 \beta ^2 & 0 & 0 & 0 & 0 \\
 0 & 0 & 0 & 0 & 0 & 0 & 0 & 0 & 0 & 0 & 0 & 0 & 0 & 0 & 0 & 0 \\
 0 & 0 & 0 & 0 & 0 & 0 & 0 & 0 & 0 & 0 & 0 & 0 & 0 & 0 & 0 & 0 \\
 0 & 0 & 0 & 0 & 0 & 0 & 0 & 0 & 0 & 0 & 0 & 0 & 0 & 0 & 0 & 0 \\
 0 & 0 & 0 & 0 & 0 & 0 & 0 & 0 & 0 & 0 & 0 & 0 & 0 & 0 & 0 & 0 \\
\end{array}
\right),
\end{equation}
\end{widetext}where we have used the same replacements as that of the first case, i.e. $\sin \left(r_c\right)\to \alpha , \sin \left(r_d\right)\to \beta , \cos \left(r_c\right)\to \gamma , \cos \left(r_d\right)\to \delta$.

\begin{widetext}
\begin{equation}\label{}
\rho_{ABC_{I}D_{I}}^{T_{A}}=\frac{1}{4}\left(
\begin{array}{cccccccccccccccc}
 0 & 0 & 0 & 0 & 0 & 0 & 0 & 0 & 0 & \gamma ^2 \delta  & \gamma  \delta ^2 & 0 & \gamma ^2 \delta ^2 & 0 & 0 & 0 \\
 0 & \gamma ^2 & \gamma  \delta  & 0 & \gamma ^2 \delta  & 0 & 0 & 0 & 0 & 0 & 0 & \beta ^2 \gamma  & 0 & \beta ^2 \gamma ^2 & 0 & 0 \\
 0 & \gamma  \delta  & \delta ^2 & 0 & \gamma  \delta ^2 & 0 & 0 & 0 & 0 & 0 & 0 & \alpha ^2 \delta  & 0 & 0 & \alpha ^2 \delta ^2 & 0 \\
 0 & 0 & 0 & \alpha ^2+\beta ^2 & 0 & \beta ^2 \gamma  & \alpha ^2 \delta  & 0 & 0 & 0 & 0 & 0 & 0 & 0 & 0 & \alpha ^2 \beta ^2 \\
 0 & \gamma ^2 \delta  & \gamma  \delta ^2 & 0 & \gamma ^2 \delta ^2 & 0 & 0 & 0 & 0 & 0 & 0 & 0 & 0 & 0 & 0 & 0 \\
 0 & 0 & 0 & \beta ^2 \gamma  & 0 & \beta ^2 \gamma ^2 & 0 & 0 & 0 & 0 & 0 & 0 & 0 & 0 & 0 & 0 \\
 0 & 0 & 0 & \alpha ^2 \delta  & 0 & 0 & \alpha ^2 \delta ^2 & 0 & 0 & 0 & 0 & 0 & 0 & 0 & 0 & 0 \\
 0 & 0 & 0 & 0 & 0 & 0 & 0 & \alpha ^2 \beta ^2 & 0 & 0 & 0 & 0 & 0 & 0 & 0 & 0 \\
 0 & 0 & 0 & 0 & 0 & 0 & 0 & 0 & \gamma ^2 \delta ^2 & 0 & 0 & 0 & 0 & 0 & 0 & 0 \\
 \gamma ^2 \delta  & 0 & 0 & 0 & 0 & 0 & 0 & 0 & 0 & \beta ^2 \gamma ^2 & 0 & 0 & 0 & 0 & 0 & 0 \\
 \gamma  \delta ^2 & 0 & 0 & 0 & 0 & 0 & 0 & 0 & 0 & 0 & \alpha ^2 \delta ^2 & 0 & 0 & 0 & 0 & 0 \\
 0 & \beta ^2 \gamma  & \alpha ^2 \delta  & 0 & 0 & 0 & 0 & 0 & 0 & 0 & 0 & \alpha ^2 \beta ^2 & 0 & 0 & 0 & 0 \\
 \gamma ^2 \delta ^2 & 0 & 0 & 0 & 0 & 0 & 0 & 0 & 0 & 0 & 0 & 0 & 0 & 0 & 0 & 0 \\
 0 & \beta ^2 \gamma ^2 & 0 & 0 & 0 & 0 & 0 & 0 & 0 & 0 & 0 & 0 & 0 & 0 & 0 & 0 \\
 0 & 0 & \alpha ^2 \delta ^2 & 0 & 0 & 0 & 0 & 0 & 0 & 0 & 0 & 0 & 0 & 0 & 0 & 0 \\
 0 & 0 & 0 & \alpha ^2 \beta ^2 & 0 & 0 & 0 & 0 & 0 & 0 & 0 & 0 & 0 & 0 & 0 & 0 \\
\end{array}
\right).
\end{equation}	
\end{widetext}
	
\begin{widetext}
\begin{equation}\label{}
\rho_{ABC_{I}D_{I}}^{T_{B}}=\frac{1}{4}\left(
\begin{array}{cccccccccccccccc}
 0 & 0 & 0 & 0 & 0 & \gamma ^2 \delta  & \gamma  \delta ^2 & 0 & 0 & 0 & 0 & 0 & \gamma ^2 \delta ^2 & 0 & 0 & 0 \\
 0 & \gamma ^2 & \gamma  \delta  & 0 & 0 & 0 & 0 & \beta ^2 \gamma  & \gamma ^2 \delta  & 0 & 0 & 0 & 0 & \beta ^2 \gamma ^2 & 0 & 0 \\
 0 & \gamma  \delta  & \delta ^2 & 0 & 0 & 0 & 0 & \alpha ^2 \delta  & \gamma  \delta ^2 & 0 & 0 & 0 & 0 & 0 & \alpha ^2 \delta ^2 & 0 \\
 0 & 0 & 0 & \alpha ^2+\beta ^2 & 0 & 0 & 0 & 0 & 0 & \beta ^2 \gamma  & \alpha ^2 \delta  & 0 & 0 & 0 & 0 & \alpha ^2 \beta ^2 \\
 0 & 0 & 0 & 0 & \gamma ^2 \delta ^2 & 0 & 0 & 0 & 0 & 0 & 0 & 0 & 0 & 0 & 0 & 0 \\
 \gamma ^2 \delta  & 0 & 0 & 0 & 0 & \beta ^2 \gamma ^2 & 0 & 0 & 0 & 0 & 0 & 0 & 0 & 0 & 0 & 0 \\
 \gamma  \delta ^2 & 0 & 0 & 0 & 0 & 0 & \alpha ^2 \delta ^2 & 0 & 0 & 0 & 0 & 0 & 0 & 0 & 0 & 0 \\
 0 & \beta ^2 \gamma  & \alpha ^2 \delta  & 0 & 0 & 0 & 0 & \alpha ^2 \beta ^2 & 0 & 0 & 0 & 0 & 0 & 0 & 0 & 0 \\
 0 & \gamma ^2 \delta  & \gamma  \delta ^2 & 0 & 0 & 0 & 0 & 0 & \gamma ^2 \delta ^2 & 0 & 0 & 0 & 0 & 0 & 0 & 0 \\
 0 & 0 & 0 & \beta ^2 \gamma  & 0 & 0 & 0 & 0 & 0 & \beta ^2 \gamma ^2 & 0 & 0 & 0 & 0 & 0 & 0 \\
 0 & 0 & 0 & \alpha ^2 \delta  & 0 & 0 & 0 & 0 & 0 & 0 & \alpha ^2 \delta ^2 & 0 & 0 & 0 & 0 & 0 \\
 0 & 0 & 0 & 0 & 0 & 0 & 0 & 0 & 0 & 0 & 0 & \alpha ^2 \beta ^2 & 0 & 0 & 0 & 0 \\
 \gamma ^2 \delta ^2 & 0 & 0 & 0 & 0 & 0 & 0 & 0 & 0 & 0 & 0 & 0 & 0 & 0 & 0 & 0 \\
 0 & \beta ^2 \gamma ^2 & 0 & 0 & 0 & 0 & 0 & 0 & 0 & 0 & 0 & 0 & 0 & 0 & 0 & 0 \\
 0 & 0 & \alpha ^2 \delta ^2 & 0 & 0 & 0 & 0 & 0 & 0 & 0 & 0 & 0 & 0 & 0 & 0 & 0 \\
 0 & 0 & 0 & \alpha ^2 \beta ^2 & 0 & 0 & 0 & 0 & 0 & 0 & 0 & 0 & 0 & 0 & 0 & 0 \\
\end{array}
\right).
\end{equation}	
\end{widetext}

\begin{widetext}
\begin{equation}\label{}
\rho_{ABC_{I}D_{I}}^{T_{C_{I}}}=\frac{1}{4}\left(
\begin{array}{cccccccccccccccc}
 0 & 0 & 0 & \gamma  \delta  & 0 & 0 & \gamma  \delta ^2 & 0 & 0 & 0 & \gamma  \delta ^2 & 0 & 0 & 0 & 0 & 0 \\
 0 & \gamma ^2 & 0 & 0 & \gamma ^2 \delta  & 0 & 0 & \beta ^2 \gamma  & \gamma ^2 \delta  & 0 & 0 & \beta ^2 \gamma  & 0 & 0 & 0 & 0 \\
 0 & 0 & \delta ^2 & 0 & 0 & 0 & 0 & 0 & 0 & 0 & 0 & 0 & 0 & 0 & 0 & 0 \\
 \gamma  \delta  & 0 & 0 & \alpha ^2+\beta ^2 & 0 & 0 & \alpha ^2 \delta  & 0 & 0 & 0 & \alpha ^2 \delta  & 0 & 0 & 0 & 0 & 0 \\
 0 & \gamma ^2 \delta  & 0 & 0 & \gamma ^2 \delta ^2 & 0 & 0 & 0 & \gamma ^2 \delta ^2 & 0 & 0 & 0 & 0 & 0 & 0 & 0 \\
 0 & 0 & 0 & 0 & 0 & \beta ^2 \gamma ^2 & 0 & 0 & 0 & \beta ^2 \gamma ^2 & 0 & 0 & 0 & 0 & 0 & 0 \\
 \gamma  \delta ^2 & 0 & 0 & \alpha ^2 \delta  & 0 & 0 & \alpha ^2 \delta ^2 & 0 & 0 & 0 & \alpha ^2 \delta ^2 & 0 & 0 & 0 & 0 & 0 \\
 0 & \beta ^2 \gamma  & 0 & 0 & 0 & 0 & 0 & \alpha ^2 \beta ^2 & 0 & 0 & 0 & \alpha ^2 \beta ^2 & 0 & 0 & 0 & 0 \\
 0 & \gamma ^2 \delta  & 0 & 0 & \gamma ^2 \delta ^2 & 0 & 0 & 0 & \gamma ^2 \delta ^2 & 0 & 0 & 0 & 0 & 0 & 0 & 0 \\
 0 & 0 & 0 & 0 & 0 & \beta ^2 \gamma ^2 & 0 & 0 & 0 & \beta ^2 \gamma ^2 & 0 & 0 & 0 & 0 & 0 & 0 \\
 \gamma  \delta ^2 & 0 & 0 & \alpha ^2 \delta  & 0 & 0 & \alpha ^2 \delta ^2 & 0 & 0 & 0 & \alpha ^2 \delta ^2 & 0 & 0 & 0 & 0 & 0 \\
 0 & \beta ^2 \gamma  & 0 & 0 & 0 & 0 & 0 & \alpha ^2 \beta ^2 & 0 & 0 & 0 & \alpha ^2 \beta ^2 & 0 & 0 & 0 & 0 \\
 0 & 0 & 0 & 0 & 0 & 0 & 0 & 0 & 0 & 0 & 0 & 0 & 0 & 0 & 0 & 0 \\
 0 & 0 & 0 & 0 & 0 & 0 & 0 & 0 & 0 & 0 & 0 & 0 & 0 & 0 & 0 & 0 \\
 0 & 0 & 0 & 0 & 0 & 0 & 0 & 0 & 0 & 0 & 0 & 0 & 0 & 0 & 0 & 0 \\
 0 & 0 & 0 & 0 & 0 & 0 & 0 & 0 & 0 & 0 & 0 & 0 & 0 & 0 & 0 & 0 \\
\end{array}
\right).
\end{equation}	
\end{widetext}

\begin{widetext}
\begin{equation}\label{}
\rho_{ABC_{I}D_{I}}^{T_{D_{I}}}=\frac{1}{4}\left(
\begin{array}{cccccccccccccccc}
 0 & 0 & 0 & \gamma  \delta  & 0 & \gamma ^2 \delta  & 0 & 0 & 0 & \gamma ^2 \delta  & 0 & 0 & 0 & 0 & 0 & 0 \\
 0 & \gamma ^2 & 0 & 0 & 0 & 0 & 0 & 0 & 0 & 0 & 0 & 0 & 0 & 0 & 0 & 0 \\
 0 & 0 & \delta ^2 & 0 & \gamma  \delta ^2 & 0 & 0 & \alpha ^2 \delta  & \gamma  \delta ^2 & 0 & 0 & \alpha ^2 \delta  & 0 & 0 & 0 & 0 \\
 \gamma  \delta  & 0 & 0 & \alpha ^2+\beta ^2 & 0 & \beta ^2 \gamma  & 0 & 0 & 0 & \beta ^2 \gamma  & 0 & 0 & 0 & 0 & 0 & 0 \\
 0 & 0 & \gamma  \delta ^2 & 0 & \gamma ^2 \delta ^2 & 0 & 0 & 0 & \gamma ^2 \delta ^2 & 0 & 0 & 0 & 0 & 0 & 0 & 0 \\
 \gamma ^2 \delta  & 0 & 0 & \beta ^2 \gamma  & 0 & \beta ^2 \gamma ^2 & 0 & 0 & 0 & \beta ^2 \gamma ^2 & 0 & 0 & 0 & 0 & 0 & 0 \\
 0 & 0 & 0 & 0 & 0 & 0 & \alpha ^2 \delta ^2 & 0 & 0 & 0 & \alpha ^2 \delta ^2 & 0 & 0 & 0 & 0 & 0 \\
 0 & 0 & \alpha ^2 \delta  & 0 & 0 & 0 & 0 & \alpha ^2 \beta ^2 & 0 & 0 & 0 & \alpha ^2 \beta ^2 & 0 & 0 & 0 & 0 \\
 0 & 0 & \gamma  \delta ^2 & 0 & \gamma ^2 \delta ^2 & 0 & 0 & 0 & \gamma ^2 \delta ^2 & 0 & 0 & 0 & 0 & 0 & 0 & 0 \\
 \gamma ^2 \delta  & 0 & 0 & \beta ^2 \gamma  & 0 & \beta ^2 \gamma ^2 & 0 & 0 & 0 & \beta ^2 \gamma ^2 & 0 & 0 & 0 & 0 & 0 & 0 \\
 0 & 0 & 0 & 0 & 0 & 0 & \alpha ^2 \delta ^2 & 0 & 0 & 0 & \alpha ^2 \delta ^2 & 0 & 0 & 0 & 0 & 0 \\
 0 & 0 & \alpha ^2 \delta  & 0 & 0 & 0 & 0 & \alpha ^2 \beta ^2 & 0 & 0 & 0 & \alpha ^2 \beta ^2 & 0 & 0 & 0 & 0 \\
 0 & 0 & 0 & 0 & 0 & 0 & 0 & 0 & 0 & 0 & 0 & 0 & 0 & 0 & 0 & 0 \\
 0 & 0 & 0 & 0 & 0 & 0 & 0 & 0 & 0 & 0 & 0 & 0 & 0 & 0 & 0 & 0 \\
 0 & 0 & 0 & 0 & 0 & 0 & 0 & 0 & 0 & 0 & 0 & 0 & 0 & 0 & 0 & 0 \\
 0 & 0 & 0 & 0 & 0 & 0 & 0 & 0 & 0 & 0 & 0 & 0 & 0 & 0 & 0 & 0 \\
\end{array}
\right).
\end{equation}	
\end{widetext}

For the 1-3 entanglement once again, we find out there exists some symmetry, i.e., $N_{A(BC_{I}D_{I})}$ = $N_{B(AC_{I}D_{I})}$. In the case of $N_{C_{I}(ABD_{I})}$ and $N_{D_{I}(ABC_{I})}$ it does not happen the same. (FIG.4)

On the other hand, in the 1-1 tangle, we can observe in a more particular sense the entanglement between two particles, that as we can see the maximum entanglement archived is when both particles are stationary, analogous the lower entanglement is when both accelerated particles are measured (FIG.5), where we rename the sets that based on symmetry the $1-1$ tangle can be expressed in combinations of pairs with the following sets $\kappa=\{A,B\}$, and $\xi=\{C, D\}$
\begin{eqnarray}
N_{\kappa\xi}=N_{AB}=\frac{1}{2} \left(\sqrt{2}-1\right),
\end{eqnarray}

\begin{equation}
\begin{array}{l}
N_{C_I D_I}=\frac{1}{16} \Big[\sqrt{2}  \sqrt{28 \cos  \left(2 r_c\right)
+9 \cos  \left(4 r_c\right)+27} \\
~~~~~-2 \left(\cos  \left(2 r_c\right)\right)-6\Big].
\end{array}
\end{equation} 

\begin{figure}[htbp]	
\subfigure[]{\includegraphics[width=6cm]{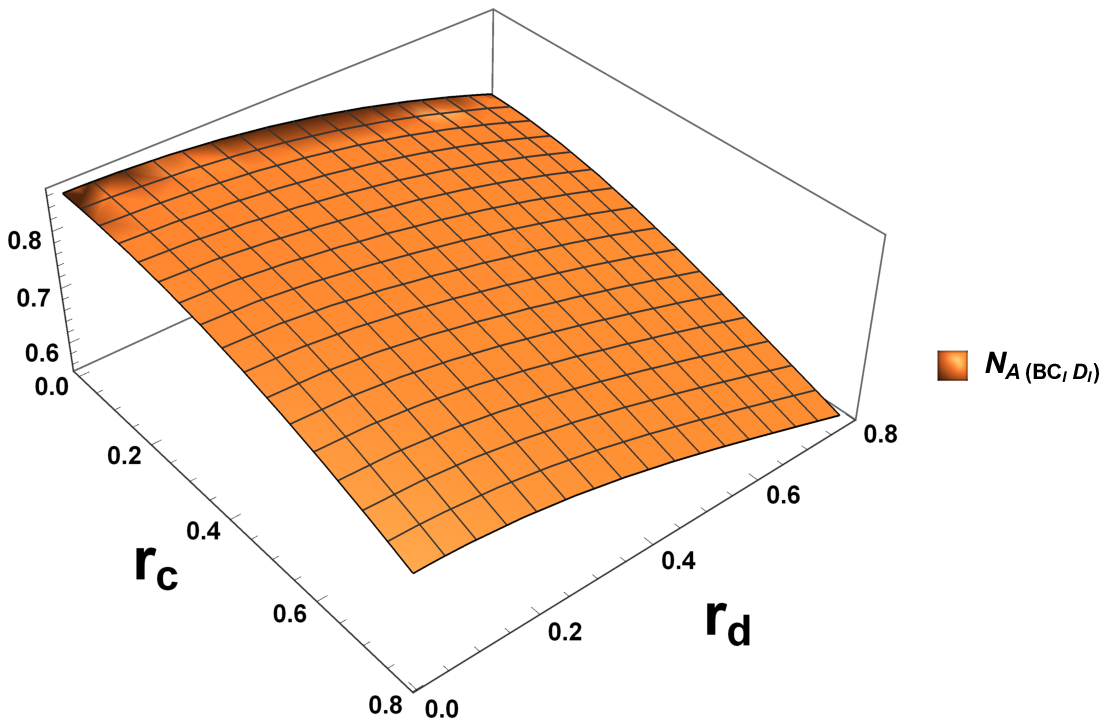}}\label{NA(BCIDI)}\hfil%
\subfigure[]{\includegraphics[width=6cm]{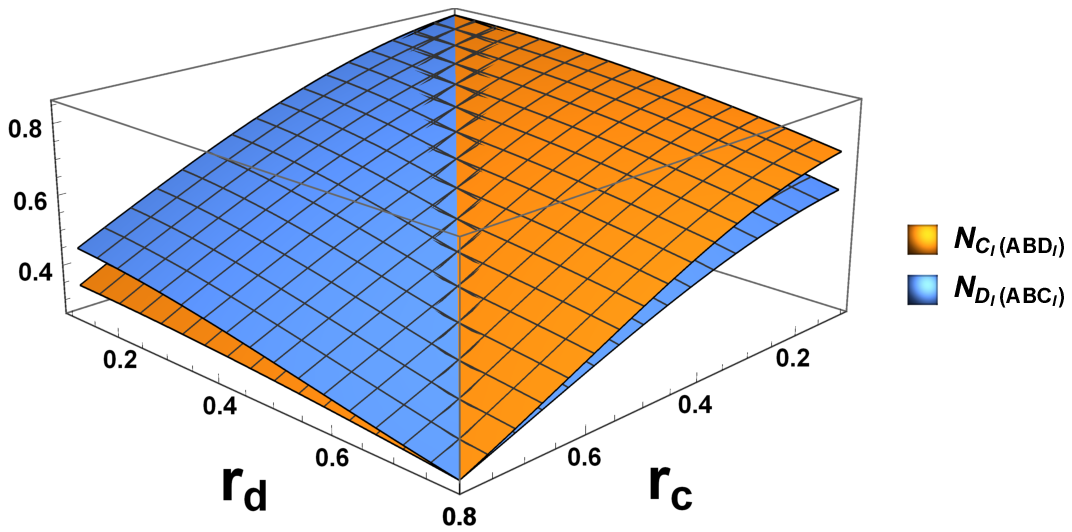}}\label{NCI(ABDI)}
\caption{\label{Negativities for two accelerated qubits} (a) Entanglement measure (1-3 tangle) from the viewpoint of Alice (Bob) as functions of acceleration parameters $r_c$ and $r_d$, (b)Entanglement measure (1-3 tangle) from the viewpoint of Charlie and David as functions of acceleration parameters $r_c$ and $r_d$.}
\end{figure}

\begin{equation}
\begin{array}{l}
N_{\kappa_{I}\xi_{I}}=\frac{1}{16} \Big[\sqrt{2} \{22 \cos  \left(2 r_c+2 r_d\right)+22 \cos  \left(2 r_c-2 r_d\right)\\[2mm]
~~~~~~~~+ 9 \cos  \left(4 r_c\right)-16 \left(\cos  \left(2 r_c\right)\right)\\[2mm]
~~~~~~~~+9 \cos  \left(4 r_d\right)-16 \left(\cos  \left(2 r_d\right)\right)+34\}^{1/2}\\[2mm]
~~~~~~~~-2 \left(\cos  \left(2 r_c+2 r_d\right)\right)-2 \left(\cos  \left(2 r_c-2 r_d\right)\right)\\[2mm]
~~~~~~~~+2 \cos  \left(2 r_c\right)+2 \cos  \left(2 r_d\right)-8\Big].
\end{array}
\end{equation}

The quantity $N_{\kappa D_{I}}$ can be obtained easily by replacing $r_{c}$ in $N_{\kappa\xi_{I}}$ with $r_{d}$.


\begin{figure}[htbp]{\includegraphics[width=5cm]{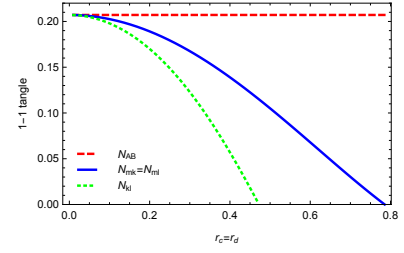}}\label{1-1 tangle}
\caption{\label{Residual entanglement}\text{Entanglement measure (1-1 tangle) as a function of $r_c=r_d$.}}
\end{figure}

\subsection{The $\pi$-tangle entanglement measure}

Again, using the method required for a tetrapartite state \cite{Oliveira} , we will use equations (\ref{eq24-27})-(27) to obtain the residual entanglement for the W-class state. For this case, we have
\begin{equation}
\begin{split}
\pi_{A}=N_{A(BC_ID_I)}^{2}-N_{AB}^{2}-N_{AC_I}^{2}-N_{AD_I}^2,\\
\pi_{B}=N_{B(AC_ID_I)}^{2}-N_{BA}^{2}-N_{BC_I}^{2}-N_{BD_I}^2,\\
\pi_{C_I}=N_{C_I(ABD_I)}^{2}-N_{C_IA}^{2}-N_{C_IB}^{2}-N_{C_ID_I}^2,\\
\pi_{D_I}=N_{D_I(ABC_I)}^{2}-N_{D_IA}^{2}-N_{D_IB}^{2}-N_{D_IC_I}^2,
\end{split}
\end{equation}from which we have
\begin{eqnarray}
\pi_{4}=\frac{1}{4}\left(\pi_A+\pi_B+\pi_{C_I}+\pi_{D_I}\right)
\end{eqnarray}
and the geometric average (\ref{eq29}):
\begin{eqnarray}
\Pi_4=\left(\pi_A\pi_B\pi_{C_I}\pi_{D_I}\right)^{\frac{1}{4}}.
\end{eqnarray}

Since the obtained calculations are not easy to express in a short term, we shall only present their representative graph in FIG.6. We remark the antisymmetry between $\pi_{C_I}$ and $\pi_{D_I}$, although $\pi_{A}$ and $\pi_{B}$ remain the same. As we know, it is very difficult to plot $\pi_{CI}$ and $\pi_{DI}$ as 3D graphics since the negativity $N_{C_{I}D_{I}}$ is equal to zero when $r>0.472473$. The graphics $\pi_{CI}$ and $\pi_{DI}$ are plotted in two segmentation intervals of the variable $r$, i.e. $r\in [0,0.472473]\bigcup [0.472473, \pi/4]$.

\begin{figure}[htbp]	
\subfigure[]{\includegraphics[width=8cm]{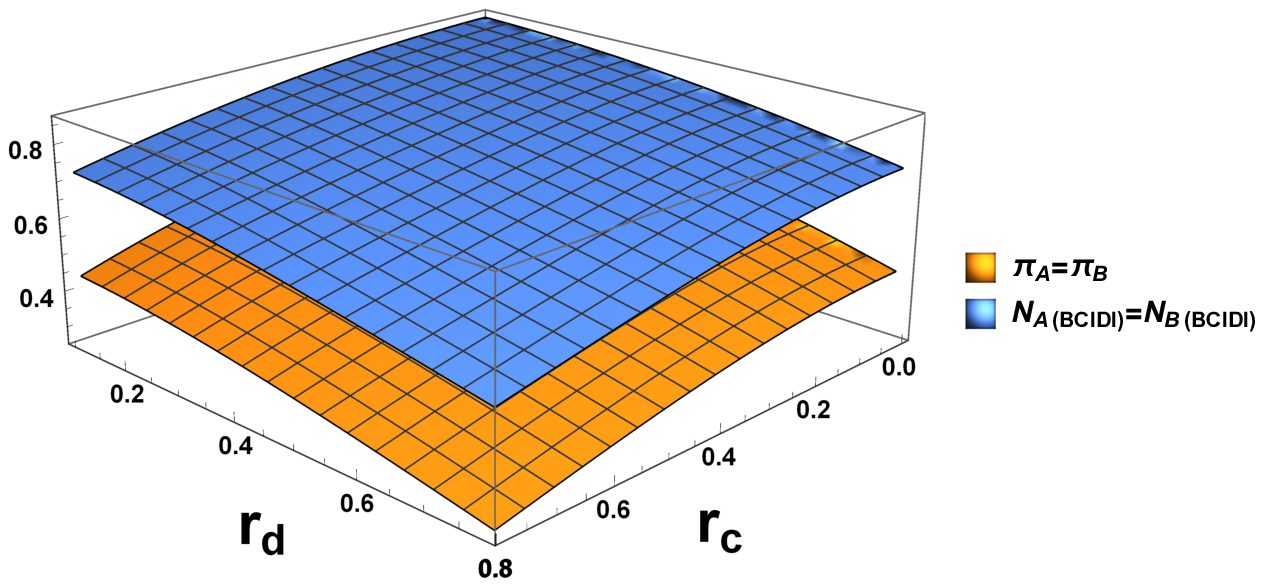}}\label{pi_cd}\hfil%
\subfigure[]{\includegraphics[width=8cm]{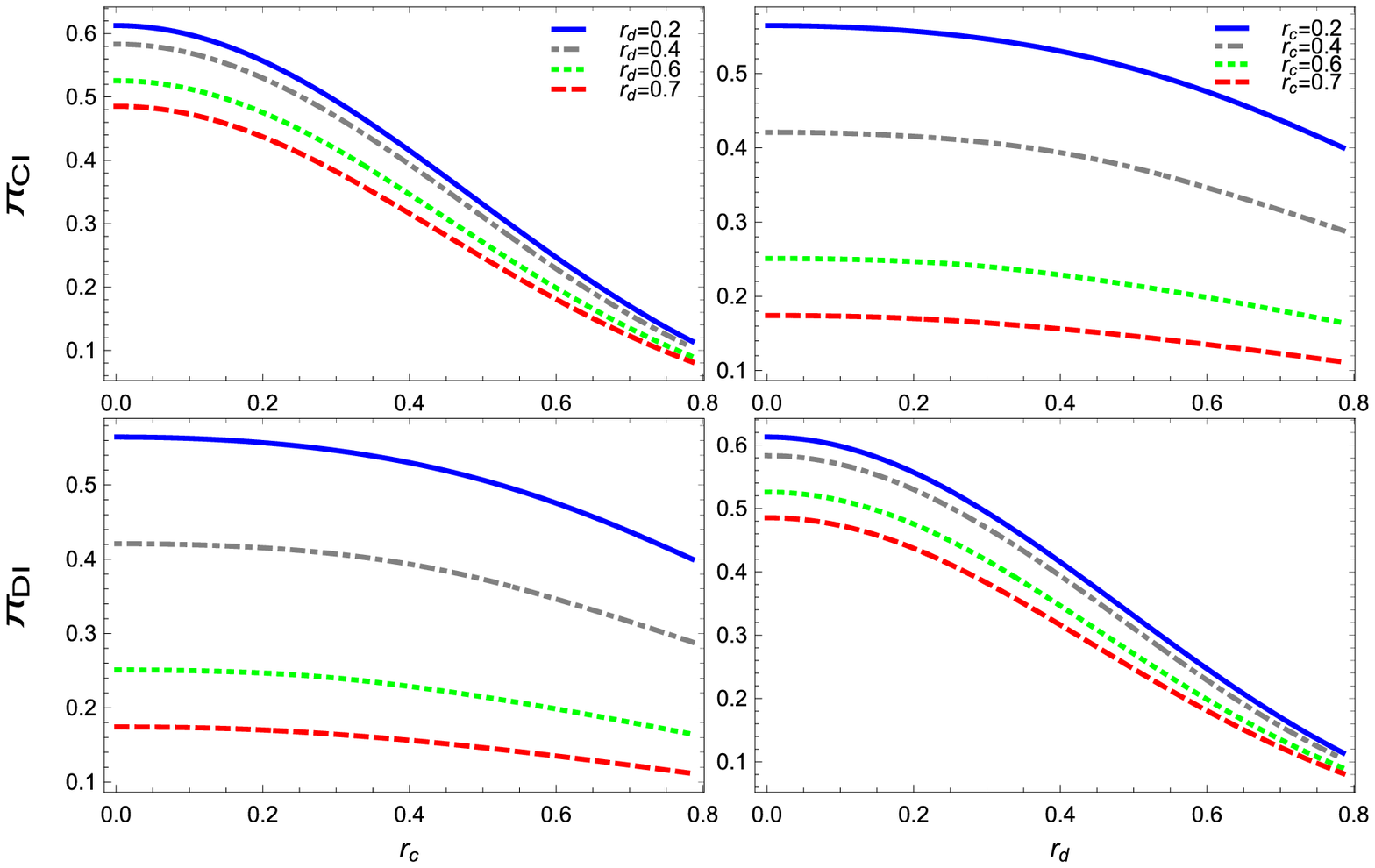}}\label{pi_a}
\caption{(a)Residual Entanglement ($\pi$-tangle) for Alice and Bob as well as their comparison with their 1-3 tangle, (b)Residual Entanglement ($\pi$-tangle) for Charlie and David.}
\end{figure}

Finally, we will present in FIG.7 the $\pi_4$-tangle and the whole entanglement calculated with equations (\ref{eq28}) and (\ref{eq29}).

\begin{figure}[htbp]	
\includegraphics[width=8cm]{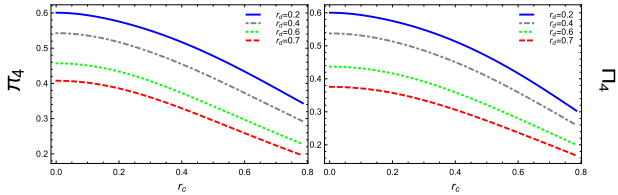}
\caption{Whole entanglement measures($\pi_4$ and $\Pi_4$) as a function of acceleration parameters $r_c$. The ($\pi_4$ and $\Pi_4$) as a function of the parameter $r_d$ are same as above.}
\end{figure}

\section{von Neumann entropy}
Another useful measurement for entanglement was inspired by von Neumann, establishing the analogous in Quantum Mechanics of the Shannon entropy for classical probability nowadays called as von Neumann entropy \cite{VoN, Popescu, Vedral}. It is defined as $S=-\mathrm{Tr}(\rho \ln \rho)$. Another definition is developed and provided by Bengtsson and \.{Z}yczkowski \cite{Bengtsson}
\begin{eqnarray}
S=-\sum_{i=1}^{N} \lambda_{i} \ln{\lambda_{i}},
\end{eqnarray}
where $\lambda_{i}$ denotes the $i$-th eigenvalue of the density matrix $\rho$.
Using this definition we are able to calculate the von Neumann entropy. Since for the case when only David is accelerated (FIG.8) all eigenvalues except for two are zero, and for the case when Charlie and David are moving, all eigenvalues except for four are zero, we are able to obtain the explicit analytical expression for the von Neumann entropy in both cases, but for the case when two qubits are accelerated, the expression will be omitted due to long-polynomial nature. Nevertheless, we show the behavior of the entropy for this case in FIG.9:
\begin{equation}
\begin{split}
S_D=\frac{3}{8} \left(\cos \left(2 r_d\right)-1\right) \ln \left(\frac{1}{8} (-3) \left(\cos \left(2 r_d\right)-1\right)\right)\\
-\frac{1}{8} \left(3 \cos \left(2 r_d\right)+5\right) \ln \left(\frac{1}{8} \left(3 \cos \left(2 r_d\right)+5\right)\right).
\end{split}
\end{equation} We find that the von Neumann entropy for these two cases increase when the acceleration is increasing.


\begin{figure}[htbp]
\includegraphics[width=7cm]{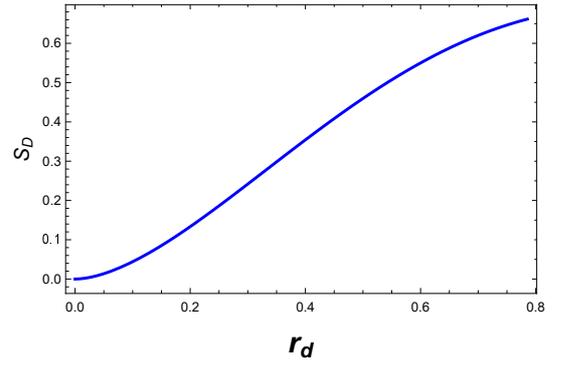}\label{von Neumann Entropy}\hfil%
\caption{ von Neumann Entropy as David's acceleration parameter $r_d$ increases.}
\end{figure}
\begin{figure}[htbp]
\includegraphics[width=7cm]{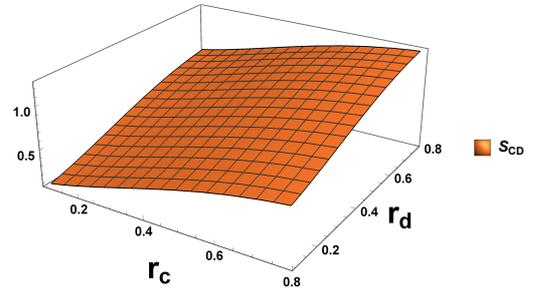}\label{von Neumann Entropy 2 moving}\hfil%
\caption{von Neumann Entropy as a function of Charlie's and David's acceleration parameters $r_c$ and $r_d$.}
\end{figure}

\section{Discussions and concluding remarks}
In conclusion, it is found that the Negativity and whole entanglement computations of the 4-particle W-class state, decrease the entanglement when we measure in a non-inertial frame. As we can see, the entanglement measures that depend on acceleration parameters will decrease over when increasing accelerations. On the contrary, the entanglement measures that do not depend on the acceleration parameter will remain a constant in acceleration limit. It is worth noting that no matter which qubit is selected from computations, we can choose any qubit or any pair of qubits from the system and switch it from inertial to a noninertial, without expecting different results in any case. For the case of the $1-1$ tangle, the entanglement begins to disappear, for the case of one accelerated qubit. We can see that when we measure the bipartite negativity ($N_{{AB}}$) never can reach infinite accelerations. The lost of entanglement is due to the 4 total partial traces that we are preforming yielding to a ripping off information. The $N_{C_ID_I}$ disappears for $r>0.472473$. This result is contrary to the cases reported for tripartite systems in which the entanglement has never been reduced to zero. For the case of the $1-3$ tangle, when we are studying the one accelerated observer case, it is shown that the entanglement measure from the point of view of the noninertial qubit decreases faster than that when the other stationary qubits preform the measure. Furthermore, when we study the second case, we can see that the negativity for the stationary qubits behaves the same over all the accelerations, but when the negativities are measured for the accelerated ones, we can see their entanglement is different. It is remarkable to say that for this measure the entanglement never reaches zero. The Whole entanglement measurements $\pi_4$ and $\Pi_4$ does not change their initial entanglement, no matter if we are studying the first or the second case, the arithmetic mean is greater than the geometric mean. Our results also suggest that, due to the growth of von Neumann entropy, our system becomes more and more mixed as we vary the acceleration parameter. Special cases were treated for the two noninertial case when $r_c$ = $r_b$, having a bigger entropy than the fist case. Before ending this work, we give a useful remark about the tetrapartite entanglement measurements for the W-class state. In noninertial frames, we will predict the difficulty of studying these entanglement measures when three fermions, say Bob, Charlie and David, are moving with nonuniform accelerations $r_{b},r_{c}$ and $r_{d}$. This is because it is impossible to illustrate them in graphics with three variables. Similarly, it will become more difficult to study the case when all particles are moving with nonuniform accelerations.

\section*{Acknowledgements}
This work was partially supported by the CONACYT, Mexico under the Grant No. 288856-CB-2016, partially by 20180677-SIP-IPN, Mexico and partially by the program XXVIII Verano de la Investigaci\'on Cient\'ifica 2018 supported by the Academia Mexicana de Ciencias.

\end{document}